# Study of induced defects of Ascorbic Acid incorporated in Graphene Oxide

Maria Fernanda Malateaux[1], Fernando Sanchez[1], Pedro Holland de Oliveira[1], Raíssa Chaves[1], Igor Abê[2], Daniel Patiño[1] and Ana Champi[1*]

[1]Laboratory of New Carbon Materials: Graphene, Center of Natural and Human Sciences, Federal University of ABC, Santo André, SP 09210-580, Brazil.

[2]Universidade de São Paulo – Escola Politécnica, São Paulo-SP. Brazil.

[*]Corresponding author: ana.champi@ufabc.edu.br

## Abstract

This work reports a study on the structural defect due to the interaction between graphene oxide, synthesized via the modified Hummers method, and L-ascorbic acid (AA). The functionalization was carried out at 65 °C, and the evolution of the material's properties was investigated as a function of time (15, 30, 45, 60, and 120 min). The resulting GO-AA samples were characterized by Fourier Transform Infrared (FTIR) spectroscopy, Raman spectroscopy, and Scanning Electron Microscopy (SEM). The analysis demonstrate that L-ascorbic acid was functionalized with GO, as evidenced by FTIR, which showed a change in the intensity of characteristic oxygenated bands, while Raman spectroscopy shows a variation in the intensity ratio between the D and G bands ($I_D/I_G$), and an increase in defect density with interaction time. SEM images reveal morphological changes from smooth GO sheets to a more compact and crumpled GO-AA structure with increasing interaction time. Overall, the results demonstrate that interaction with L-ascorbic acid induces structural defects in graphene oxide.

## 1. INTRODUCTION

Graphene has garnered significant scientific attention due to its unique structure as a two-dimensional (2D) lattice of $sp^2$-hybridized carbon atoms arranged in a hexagonal pattern. Its extraordinary properties, including mechanical strength superior to steel, high electron mobility, and excellent thermal conductivity, make it a promising material for different applications such as electronics and devices as batteries and supercapacitors [1-3]. While graphene offers numerous advantages, transitioning from laboratory synthesis to industrial-level output still remains an obstacle. The mechanical exfoliation method, where adhesive tape is used to peel graphene layers down to a single layer, while capable of producing high-quality graphene, suffers from low yield and is not scalable for mass production [4-6]. A viable, low-cost, high-yield alternative is the chemical route such as the Hummers method and the Marcano method [7, 10, 12, 21], both of which involve the oxidation of graphite to produce graphene oxide (GO).

In particular, the Hummers method utilizes strong oxidizers to insert oxygen-containing functional groups (like hydroxyls, epoxides, and carboxyl) into the graphite structure to synthesize the graphene oxide. This process increases the interplanar spacing and renders the material hydrophilic, allowing for exfoliation in water to form GO suspensions [1]. However, the introduction of these functional groups disrupts the conjugated $sp^2$ network, making GO an electrical insulator [3].

To restore the conductive properties and electronic conjugation of graphene, GO must undergo a reduction process to become reduced graphene oxide (rGO). Reduced Graphene oxide can be obtained through different routes, classified as thermal, irradiation, electrochemical, and chemical reduction methods [18]. Chemical reduction remains the most widely adopted approach due to its economic viability, potential for large-scale production, and mild thermal requirements.

Historically, reducing agents like hydrazine have been widely used in the reduction process due to their effectiveness. However, hydrazine is "highly poisonous and explosive" and presents significant risks to health and the environment. In recent years, a "green reduction" approach has gained traction, seeking non-toxic alternatives. L-ascorbic acid (AA), or Vitamin C, has emerged as an ideal substitute [2, 4, 11]. AA is considered a mild, non-toxic, and environmentally friendly reducing agent [11]. Studies demonstrate that AA is effective in removing most of the oxygen functionalities from GO, restoring the aromatic structure, and consequently forming the rGO [4, 5].

However, differently from what is reported in the literature, this work did not use hydroxides, or other chemical agents like $NH_3$, to alkalize the pH. In this regard, it is noted that in an acidic medium, the presence of ascorbic acid alone was not sufficient to induce GO reduction [3], and this is evident from the presence of oxygenated groups in the structure, which was revealed by the FTIR technique. Therefore, in this study, Ascorbic Acid (AA) is functionalized with graphene oxide (GO) to investigate the defects generated due to the AA incorporate in the GO matrix to five different times and constant temperature, in order to investigate what structure is being formed.

By FTIR, information on the oxygenated functional groups and their relationship with the defects observed by Raman through $I_D/I_G$ and the defect density ($n_D$) was observed, as well as an inversely proportional relationship with the crystallite size ($L_a$) obtained from the Raman spectra as a function of the heat treatment time with AA. Finally, SEM analysis unveiled slight differences in the surface morphology of the GO AA samples for different times.

## 2. EXPERIMENTAL PROCEDURES

### 2.1 Graphene Oxide (GO)

#### 2.1.A Materials

Natural graphite flakes (99.5% purity) were provided by Nacional de Grafite, in Brazil. Sodium nitrate ($NaNO_3$, 99%), potassium permanganate ($KMnO_4$, 99%), sulfuric acid ($H_2SO_4$, 98%), hydrogen peroxide ($H_2O_2$, 29%), hydrochloric acid (HCl, 37%), and absolute ethanol were acquired from LabSynth [23].

Ultrapure and distilled water were used for the preparation and washing of all nanostructures.

#### 2.1.B Synthesis of Graphene Oxide (GO)

Graphene oxide was synthesized using the modified Hummers method [8].

3 g of natural graphite, 1.5 g of $NaNO_3$, and 69 mL of $H_2SO_4$ were mixed under magnetic stirring for 5 minutes. The solution was then cooled to 5 °C in an ice bath. Subsequently, 9 g of $KMnO_4$ was added, ensuring the reaction temperature was maintained below 20 °C due to the exothermic nature of the reaction [23]. After the addition was complete, the temperature was raised to 35 °C and held for 30 minutes. Following this, 138 mL of ultrapure water was carefully added, and the reaction was allowed to proceed for another 15 minutes [23].

To terminate the oxidation, 420 mL of ultrapure water and 3 mL of $H_2O_2$ (29%) were added to the mixture. The solution was left under magnetic stirring for 1 hour. Finally, the resulting product was washed with distilled water, HCl, and ethanol via centrifugation until a neutral pH solution was achieved. The obtained GO was dried for subsequent use [9, 23].

## 2.2 Graphene oxide functionalized with Ascorbic acid

### 2.2.A Materials

Dried GO (Figure 1 (a)), synthesized according to the procedure in section 2.1, was used as the precursor. L-ascorbic acid ($C_6H_8O_6$) was acquired from LabSynth. Ultrapure water was used for the dispersion and reduction steps.

### 2.2.B Synthesis of Graphene Oxide-Ascorbic Acid

A GO suspension was prepared by dispersing 11.6 mg (Figure 1 (b)) of the dry GO (synthesized as described in section 2.1.B) in 100 mL of ultrapure water. The mixture was ultrasonicated for 25 minutes to ensure exfoliation, with a glass rod used to aid dissolution (Figure 1 (c)).

Following dispersion, 106 mg of ascorbic acid was added to the GO suspension after the sonication following the methodology described in [3]. The reaction mixture was then heated to 65 °C on a hot plate and maintained under constant magnetic stirring (Figure 1 (d)) [3]. To study the reduction progress, aliquots were withdrawn at specific time intervals: 15 min, 30 min, 45 min, 60 min, and 120 min (Figure 1 (e)). The resulting samples were labeled GO-AA 1, GO-AA 2, GO-AA 3, GO-AA 4, and GO-AA 5, respectively, as shown in Figure 1.

All samples were dried for 130 minutes on a hot plate, heated to 30 °C, and a solid paste was obtained.

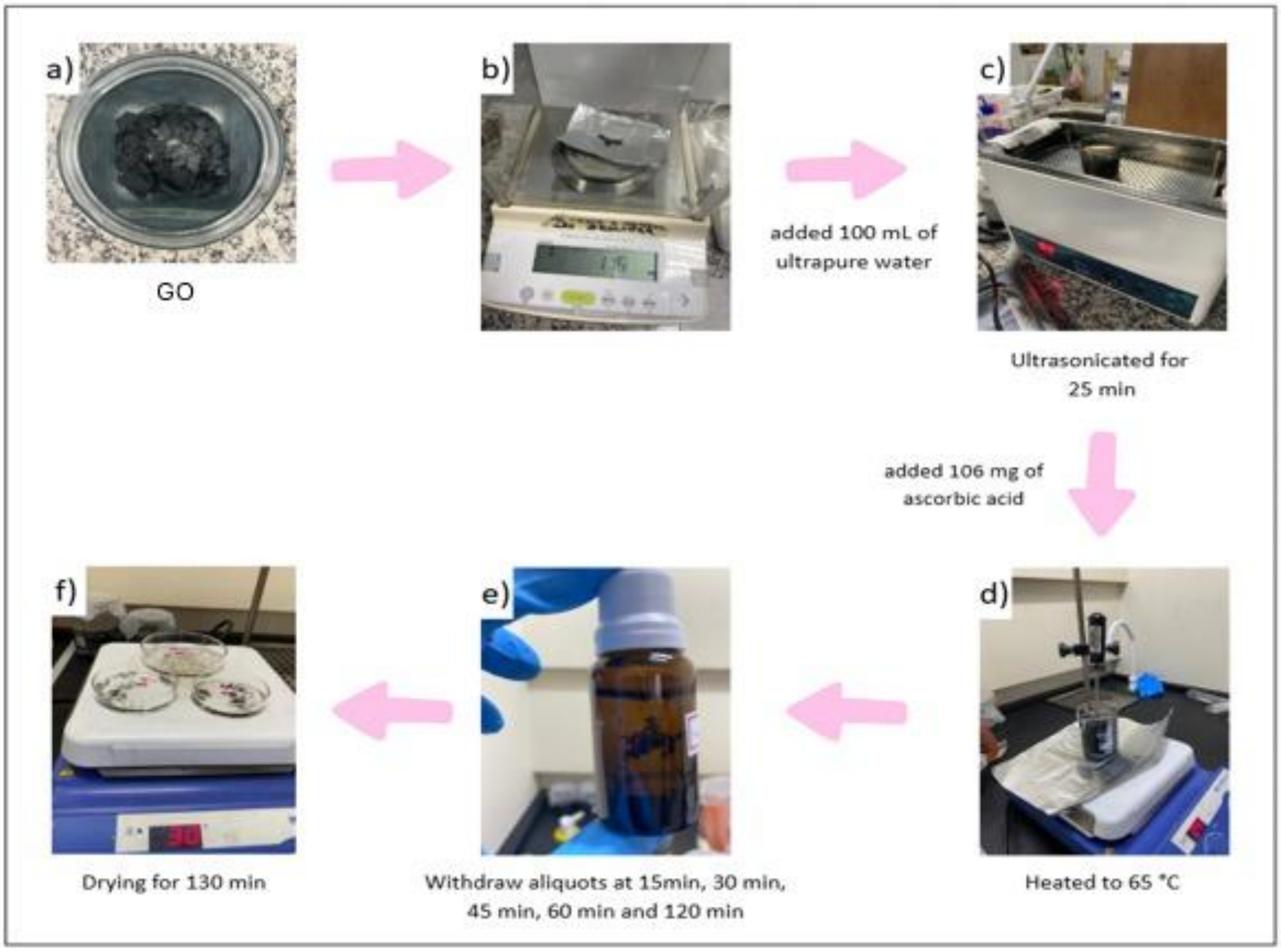

Figure 1. (a) Dried graphene oxide (GO) precursor; (b) Weighing the GO; (c) Ultrasonication of the GO dispersion in ultrapure water; (d) Heating the suspension with ascorbic acid at 65°C; (e) Withdrawing aliquots at determined time intervals of GO-AA and (f) finally GO-AA paste after drying for 130 min.

### 2.3 GO and GO-AA Characterization

The structural and morphological characterization of GO and GO-AA samples was carried out using complementary techniques. Fourier-transform infrared (FTIR) spectroscopy was conducted with a PerkinElmer Spectrum Two FT-IR Spectrometer in attenuated total reflectance (ATR) mode. Raman spectroscopy was performed with a WITec Confocal Raman Microscope alpha300 R, equipped with a green laser ($\lambda$ = 532 nm, 45 mW). The surface morphology was analyzed by scanning electron microscopy (SEM) using a JEOL JCM-6000 operated at 5 kV.

## 3. RESULTS AND DISCUSSION

### 3.1 FTIR spectroscopy

Fourier-transform infrared (FTIR) spectroscopy was employed to investigate the presence of functional groups in the samples from the obtained vibrational spectra.

Figure 2 (a) shows the full FTIR spectra of AA, GO, GO-AA 1, GO-AA 2, GO-AA 3, GO-AA 4 and GO-AA 5. This graph displays the characteristic oxygenated vibrational modes described by Brusko et al. [13], including the C=O stretching of carboxyl groups (~1720 $cm^{-1}$), the $H_2O$ bending (~1620 $cm^{-1}$), the organo-sulfate bands S=O (~1410 and ~1223 $cm^{-1}$), the C–OH bending (~1280 $cm^{-1}$), and the C–O and epoxy-related signals in the 1040–980 $cm^{-1}$ region. In addition, it shows the -OH (highlighted) band which goes from 2900-3800 $cm^{-1}$.

The spectra presented in Figure 2 (b) highlight the evolution of oxygen-containing groups in GO and GO-AAs from 900 $cm^{-1}$ to 1500 $cm^{-1}$. It shows a dominant peak associated with the C–O stretching vibration of epoxy functionalities. The peak, near 980 $cm^{-1}$, decreases significantly after functionalization with Ascorbic Acid, particularly for GO-AA 2, GO-AA 3, GO-AA 4 and GO-AA 5, indicating the effective opening of epoxide rings by ascorbic acid. In addition, the tertiary alcohol C-O stretching (1030–1050 $cm^{-1}$) also decreases progressively, demonstrating the removal of hydroxyl groups. Two new peaks emerge in the region between 1070 and 1120 $cm^{-1}$ corresponding to C-O $\gamma$-lactone and dioxan structures, which have been reported in association with ascorbic acid by Bichara et al. [19]. In samples subjected to longer interaction times (GO-AA 3, GO-AA 4 and GO-AA 5), a clear decrease in the intensity of the S=O (~1220 $cm^{-1}$) and S=O (~1410 $cm^{-1}$) vibrations is observed. This trend suggests the gradual elimination of sulfate-containing species, likely residues from precursor synthesis of GO [14].

Figure 2 (c) indicates the FTIR spectra of, GO and GO-AA samples in the 1850–1500 $cm^{-1}$ region. The band at ~1720 $cm^{-1}$, attributed to the C=O stretching of carbonyl groups, increases in intensity across all GO-AA samples. Similar behavior has been occasionally reported during GO-AA interaction, where partial removal of epoxy and hydroxyl groups exposes previously embedded carboxyl species, leading to an apparent C=O enhancement at intermediate interaction stages. This effect has been described by De Silva *et al.* [3] for short interaction times, where residual C=O groups remain detectable despite the loss of other oxygen functionalities. In contrast, in the present work, the C=O band stabilizes rather than decreases with longer interaction times. This behavior can be attributed to the functionalization of GO-AA, as ascorbic acid itself contains prominent carbonyl (C=O) groups, which may contribute to the general spectral intensity in this region [19].

On the other hand, the evolution of the band at 1560–1620 $cm^{-1}$ follows the expected behavior for AA-treated GO. GO exhibits a broad contribution combining $H_2O$ bending (~1620 $cm^{-1}$) and weak C=C stretching (~1580 $cm^{-1}$). Upon interaction with ascorbic acid, samples GO-AA 1 and GO-AA 2 show a decrease in intensity due to the removal of interlayer water. From GO-AA 3 onward, however, a marked increase is observed, which we attribute to the restoration of the $sp^2$ carbon network. This trend becomes more pronounced in GO-AA 4 and GO-AA 5, consistent with literature describing the strengthening of the C=C vibrational mode as the degree of graphitization increases [13].

The methodology described by Guerrero-Contreras [20] was adapted to calculate the total area associated with the oxygen functional groups. The Oxygen Related Band: Total ratio (ORB) was estimated between 900 $cm^{-1}$ and 1850 $cm^{-1}$, as shown in equation 1. The corresponding ORB values are presented as a bar chart in Figure 2d. The GO sample exhibits the lowest relative contribution of oxygen-related functional groups, whereas GO-AA 1 shows the highest oxygen content. After 20 minutes of interaction (GO-AA 2), a slight decrease in oxygen functional groups is observed, followed by a stabilization trend in samples GO-AA 3, GO-AA 4, and GO-AA 5, this demonstrates that there was no graphene oxide reduction.

$$ORB: total\ ratio = \frac{A_{total} - A_{C=C}}{A_{total}} \quad (1)$$

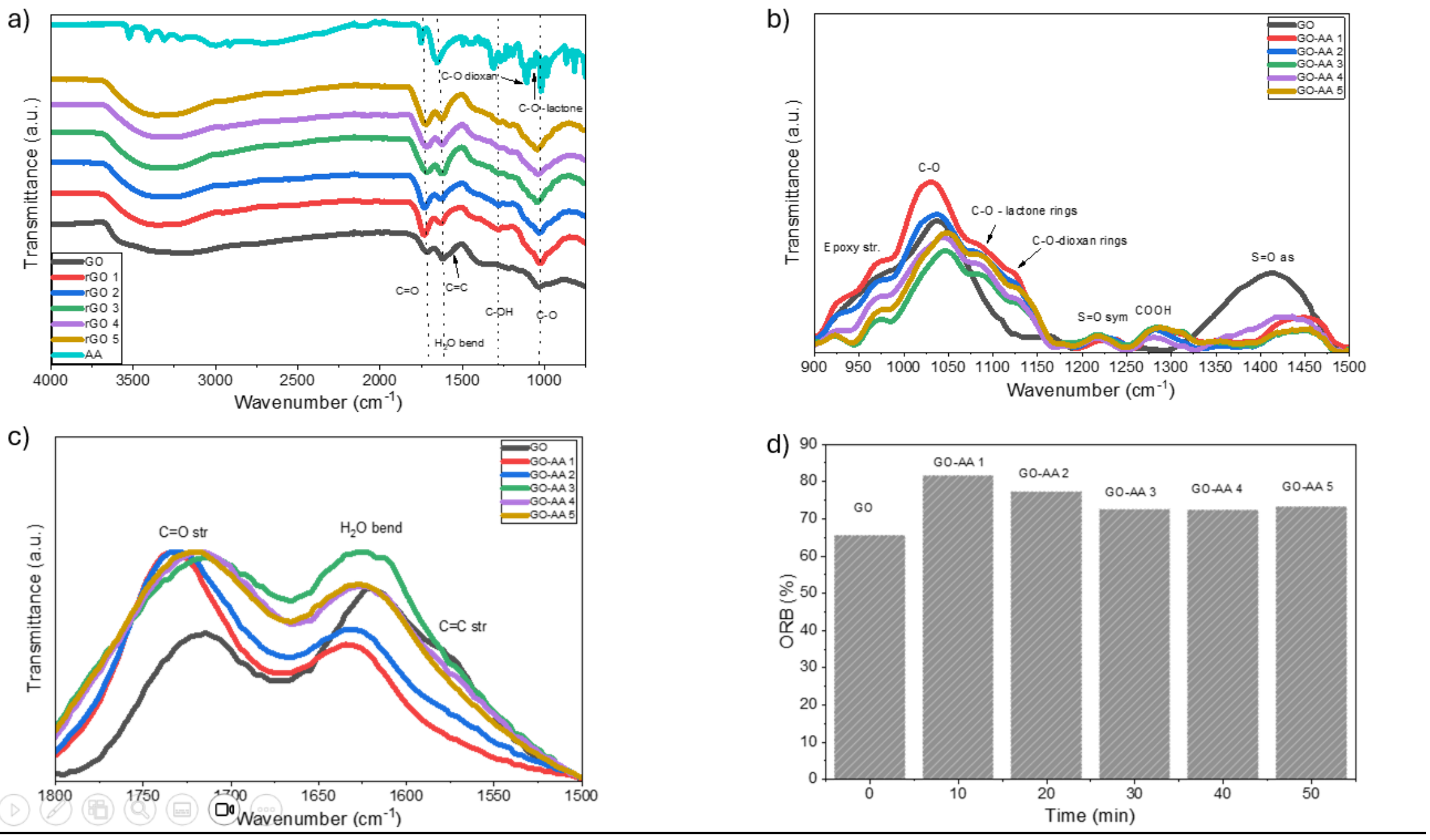

Figure 2. (a) FTIR spectrum of AA, GO and GO-AA samples according to the reduction time (15, 30, 45, 60, and 120 min); (b) vibrational modes in the region 900 until 1500cm$^{-1}$; (c) vibrational modes in the region 1500 until 1800 cm$^{-1}$ to all samples and (d) ORB: total ratio for each sample.

### 3.2 Raman Spectroscopy

Raman spectroscopy was employed to evaluate the structural evolution of graphene oxide (GO) during its functionalization with L-ascorbic acid. This technique is particularly sensitive to disorder, defect density, and the development of sp² carbon domains in graphitic materials.

The Raman spectrum is dominated by two bands in carbon materials. There is the D band (~1350 $cm^{-1}$), which is activated by structural defects, disorder, $sp^3$-hybridized carbon atoms, vacancies, and graphitic edges, emerging when the symmetry of the $sp^2$ network is disrupted. And there is the G band (1580–1600 $cm^{-1}$), which corresponds to the $E_{2g}$ vibrational mode, is associated with in-plane C=C stretching in $sp^2$-hybridized carbon systems [15].

To accurately distinguish between different vibrational modes and assess the degree of structural disorder, the raw spectral data were obtained three times for each sample in different locations. Then, the mean spectra value was obtained and subjected to a rigorous deconvolution process. The mathematical analysis was performed using a peak-fitting procedure based on five distinct Lorentzian functions, which allowed for the isolation of the overlapping D, G, D′, D″, and D* bands. The G band (~1583 $cm^{-1}$, red) corresponds to the $E_{2g}$ phonon mode of $sp^2$ carbon atoms. The D band (1330–1350 $cm^{-1}$, blue) arises from $sp^3$-type defects and lattice disorder. The D′ band (~1620 $cm^{-1}$, magenta) is attributed to intravalley scattering at impurity or edge sites. The D″ band (1500–1550 $cm^{-1}$, green) is associated with amorphous carbon and its intensity is inversely proportional to crystallinity. The D* band (1050–1200 $cm^{-1}$, orange) is characteristic of highly disordered $sp^3$-rich phases [16, 17]. Figure 3 demonstrates the contribution of each deconvoluted component to the overall signal, highlighting the evolution of defect densities and graphitic domains throughout the reaction.

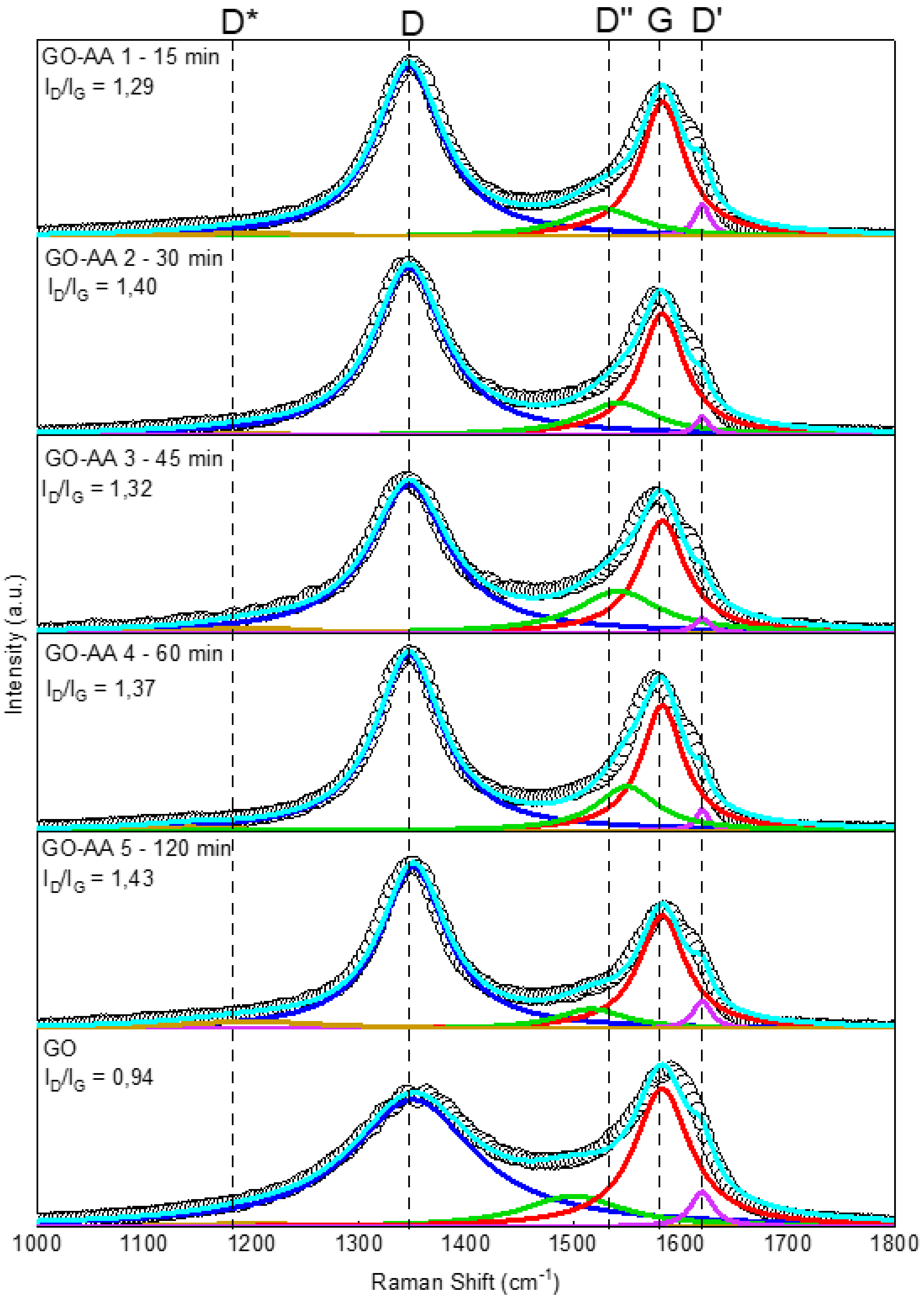

Figure 3. Raman spectra of the characteristic typical bands of GO and GOs functionalized with AA with reaction time (15, 30, 45, 60, 120 min).

GO exhibits a broad spectrum dominated by a strong D″ band, confirming a high degree of amorphization resulting from oxidative exfoliation. In all GO-AA samples, the D′ band appears consistently as a shoulder adjacent to the G band, demonstrating the persistence of impurity-related or edge defects throughout the reaction.

A particularly important trend is observed in the D* band. In GO-AA 1 through GO-AA 2, the band remains relatively weak, indicating limited formation of $sp^3$-rich disordered phases. However, in GO-AA 5 (120 min), D* becomes broader and more intense, suggesting that prolonged reaction times induce new defects, such as vacancy clusters or local structural distortions [16].

The Raman $I_D/I_G$ intensity ratio is widely used to quantify the degree of disorder in graphitic structures [15]. The characteristic band positions (D, G, D’, D’’, and D*) and the calculated $I_D/I_G$ ratios for the initial GO and the GO-AA samples obtained at different reaction times are summarized in Table 1.

Table 1 - D, G, D’, D’’, D* Bands Positions and $I_D/I_G$ ratio of obtained GO-AA samples with reaction time (15, 30, 45, 60, 120 min)

| | Bands position ($cm^{-1}$) | | | | | |
|---|---|---|---|---|---|---|
| **Sample** | **D** | **G** | **D'** | **D"** | **D*** | **$I_D/I_G$** |
| GO | 1350.54 | 1583.37 | 1619.53 | 1498.37 | 1199.99 | 0.92 |
| GO-AA 1 | 1346.39 | 1583.37 | 1619.53 | 1526.80 | 1199.99 | 1.29 |
| GO-AA 2 | 1346.39 | 1583.37 | 1619.53 | 1543.01 | 1199.99 | 1.40 |
| GO-AA 3 | 1346.39 | 1583.37 | 1619.53 | 1543.01 | 1199.99 | 1.32 |
| GO-AA 4 | 1346.39 | 1583.37 | 1619.53 | 1551.10 | 1199.99 | 1.37 |
| GO-AA 5 | 1350.54 | 1583.37 | 1619.53 | 1518.69 | 1199.99 | 1.43 |

The experimental results demonstrate a clear structural evolution of the material as a function of reaction time. As the exposure time increases, a continuous rise in the intensity ratio of the D and G bands ($I_D/I_G$) is observed, indicating the progressive insertion of defects and/or oxygen functional groups into the graphitic lattice.

To quantitatively describe this structural disorder, it is necessary to distinguish between two key parameters: the in-plane crystallite size ($L_a$) determined by equation 2, associated with one-dimensional boundary defects, and the average distance between point defects ($L_D$) determined by equation 3. According to the relations proposed by Cançado et al. [22], these parameters can be determined for graphene-based materials using visible excitation laser energies. The crystallite size follows the Tuinstra–Koenig relation:

$$L_a(nm) = (2{,}4 \times 10^{-10})\lambda_l^4 (\frac{A_D}{A_G})^{-1} \qquad (2)$$

Additionally, the distance between point defects and the corresponding defect density ($n_D$) are given by equation 4 [22]:

$$L_D^2(nm^2) = (1{,}8 \pm 0{,}5) \times 10^{-9}\lambda_l^4 (\frac{A_D}{A_G})^{-1} \qquad (3)$$

$$n_D(cm^{-2}) = \frac{(1{,}8 \pm 0{,}5) \times 10^{22}}{\lambda_l^4}(A_D/A_G) \qquad (4)$$

where $\lambda_l$ is the excitation laser wavelength in nanometers (532 nm in this study).

The evolution of these parameters is summarized in Table 2.

Table 2 - Evolution of crystallite size and defect density across the GO-AA treatment series and GO sample.

| Crystallite size and defect density | | | |
|---|---|---|---|
| **Samples** | **La (nm)** | **$L_D$ (nm)** | **$n_D$ ($1/cm^2 x10^{11}$)** |
| **GO** | 11.38 | 12.52 | 3.80 |
| **GO-AA 1** | 8.11 | 10.57 | 5.32 |
| **GO-AA 2** | 7.48 | 10.15 | 5.78 |
| **GO-AA 3** | 7.93 | 10.45 | 5.45 |
| **GO-AA 4** | 7.64 | 10.26 | 5.65 |
| **GO-AA 5** | 7.32 | 10.04 | 5.90 |

As the $I_D/I_G$ ratio rises, a continuous rise in defect density ($n_D$) can be analyzed, and increases from $3.80 \times 10^{11}$ $cm^{-2}$ for GO to $5.90 \times 10^{11}$ $cm^{-2}$ for the most extensively treated sample (GO-AA 5). Simultaneously, the crystallite size ($L_a$) decreases from 11.38 nm to 7.32 nm, indicating the fragmentation of $sp^2$ domains into smaller regions [22].

This behavior is consistent with the Tuinstra–Koenig relation, which predicts an inverse relationship between ID/IG and $L_a$. The observed increase in $I_D/I_G$ (from approximately 0.9 to 1.4) confirms that the material remains within the low-defect-density regime. In this regime, the introduction of disorder progressively reduces domain size while largely preserving the hexagonal carbon lattice, without reaching the critical point of amorphization [22].

Figure 4 illustrates the evolution of the $I_D/I_G$ ratio and crystallite size ($L_a$) as a function of reaction time (15, 30, 45, 60, and 120 min), reinforcing the correlation between increasing disorder and decreasing structural coherence.

Furthermore, this structural evolution can also be interpreted through the defect density ($n_D$). The simultaneous decrease in $L_a$ and $L_D$, along with the increase in $n_D$, confirms that both boundary and point defects are being progressively introduced during the treatment. Ultimately, the increase in the Raman $I_D/I_G$ ratio serves not only as a qualitative indicator of structural disorder, but also as a direct physical manifestation of the growing concentration of crystallite boundaries and point defects within the material [22].

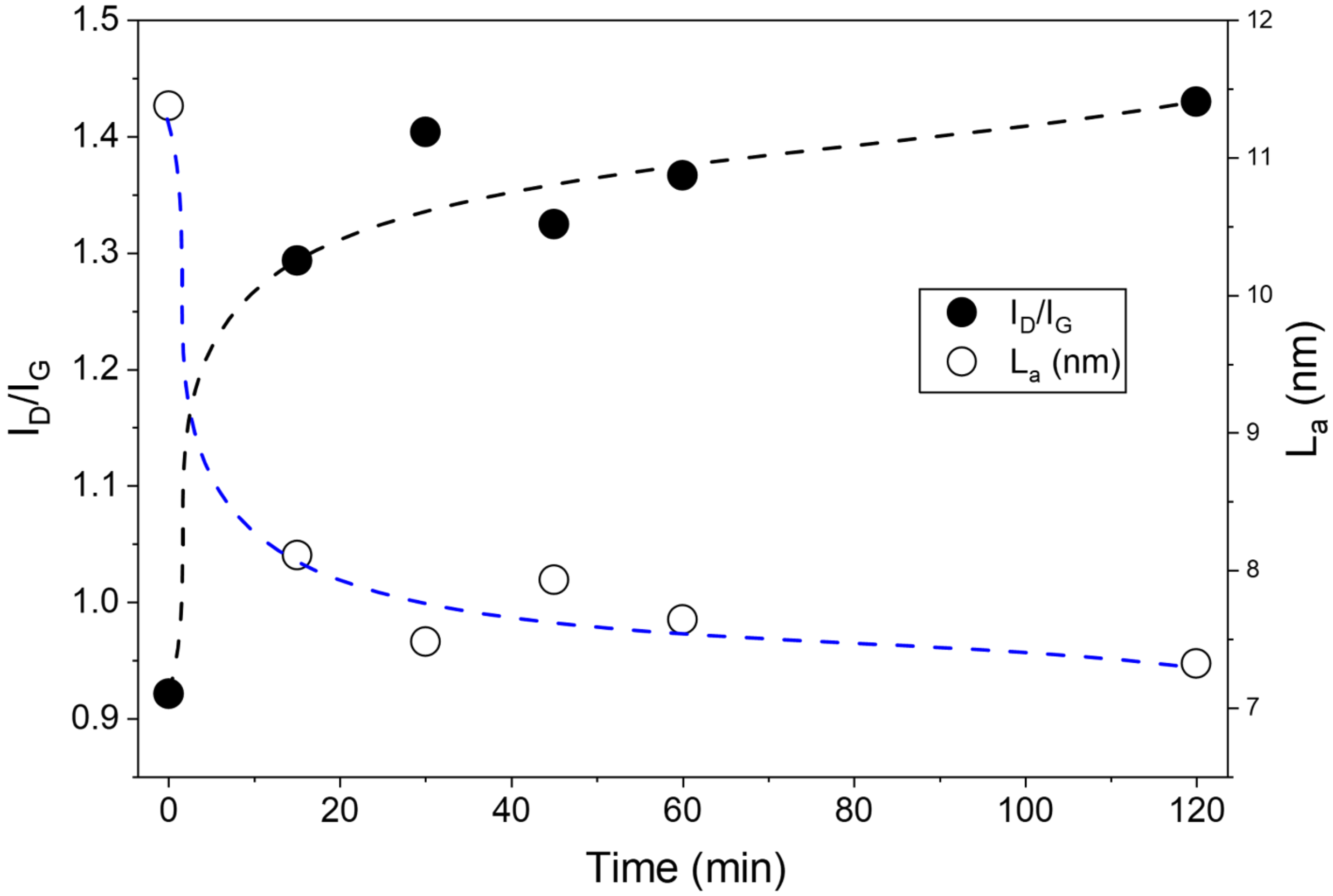

Figure 4. $I_D/I_G$ ratio and $L_a$ (nm) of obtained GO and GO-AA 1, GO-AA 2, GO-AA 3, GO-AA 4 and GO-AA 5 samples with reaction time (15, 30, 45, 60, 120 min) respectively.

### 3.3 Scanning Electron Microscopy

The morphologies of GO and GO-AA samples with different reaction times were studied by scanning electron microscopy (SEM), and the results are presented in Figure 5.

Figure 5 (a) shows the surface of graphene oxide before AA incorporation, exhibiting thin and wrinkled sheets characteristic of oxidized graphene. In contrast, the GO-AA samples (1–5) display progressively more compact and less wrinkled morphologies, resulting from the partial removal of oxygen groups and the restacking of graphene layers. Therefore, GO-AA 1 and GO-AA 2 show smoother and denser regions, GO-AA 3 and GO-AA 5 exhibit rougher and more porous surfaces, while GO-AA 4 presents a more homogeneous and compact texture.

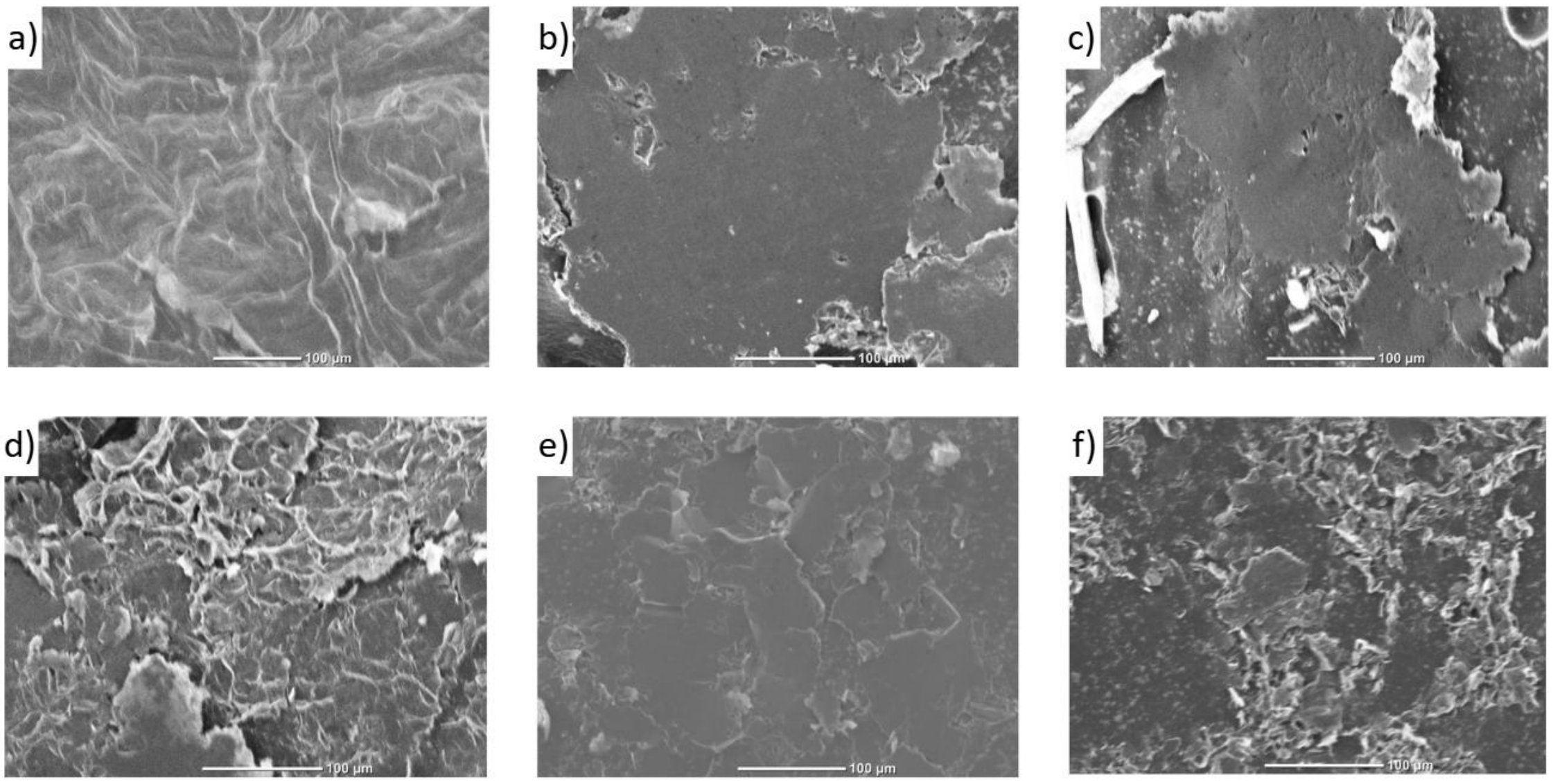

Figure 5. SEM images of (a) GO and GO-AA corresponding to reduction times (b) GO-AA 1, (c) GO-AA 2, (d) GO-AA 3, (e) GO-AA 4 and (f) GO-AA 5

## 4. CONCLUSIONS

The FTIR results reveal that the interaction between graphene oxide (GO) and L-ascorbic acid (AA) goes beyond simple deoxygenation. While AA facilitates epoxide ring-opening and the removal of specific oxygen groups, it also actively functionalizes the GO matrix. We observed the appearance of γ-lactone and dioxan rings, as well as an unexpected stabilization of carbonyl groups during longer reaction times. This shifts the perspective from a standard reduction route to a dynamic chemical functionalization process.

Structurally, Raman spectroscopy confirmed that this interaction progressively induces crystalline disorder. Instead of restoring a pristine $sp^2$ carbon network, we noted a continuous rise in the $I_D/I_G$ ratio as exposure time increased. This indicates the steady creation of structural defects, which is quantitatively supported by our calculations showing a higher point defect density ($n_D$) and a reduced in-plane crystallite size ($L_a$). At extended reaction times, localized $sp^3$-hybridized regions and vacancy clusters become increasingly prominent. Essentially, incorporating AA heavily disrupts the lattice symmetry and pushes the material into a more disordered state, although it technically remains within a low-defect-density regime.

These structural modifications are also visible morphologically. Our characterizations show a transition from the highly wrinkled sheets typical of pristine GO to a more condensed, restacked framework. The surface porosity of this new architecture varies directly with the reaction time. Gathering these observations, we demonstrate that GO AA interaction provides a highly tunable route for defect engineering. By simply adjusting the exposure time, we can deliberately control the structural disorder and defect density, offering a straightforward method to tailor graphene-based hybrids for applications that specifically rely on structural imperfections.

## 4 ACKNOWLEDGMENTS

The authors thanks for the financial support to National Council for Scientific and Technological Development – CNPq, Process: 302584/2023-4, 404412/2024-6 and FAPESP Grant (2022/11894-1). They also thank CEM at the Universidade Federal do ABC for the equipment and CAPES for the financial support.

## 5 REFERENCES

[1] Vieira Segundo, J.E.D. e Vilar, E.O. (2016). *Revista Eletrônica de Materiais e Processos*, 11(2), pp. 54-57.

[2] Faria, G.S., Lima, A.M., Brandão, L.P., Costa, A.P. da, Nardecchia, S., Ribeiro, A.A. and Pinheiro, W.A. (2018). *Matéria (Rio de Janeiro)*, 22(suppl 1). doi:https://doi.org/10.1590/s1517-707620170005.0254.

[3] De Silva, K.K.H., Huang, H.-H. and Yoshimura, M. (2018). *Applied Surface Science*, 447, pp.338–346. doi:https://doi.org/10.1016/j.apsusc.2018.03.243.

[4] Zhang, J., Yang, H., Shen, G., Cheng, P., Zhang, J. and Guo, S. (2010). *Chem. Commun.*, 46(7), pp.1112–1114. doi:https://doi.org/10.1039/b917705a.

[5] Palomba, M., Carotenuto, G. and Longo, A. (2022). *Materials*, 15(18), p.6456. doi:https://doi.org/10.3390/ma15186456.

[6] Emiru, T.F. and Ayele, D.W. (2017). *Egyptian Journal of Basic and Applied Sciences*, 4(1), pp.74–79. doi:https://doi.org/10.1016/j.ejbas.2016.11.002.

[7] Zhu, X., Liu, Q., Zhu, X., Li, C., Xu, M. and Liang, Y. (2012). *International Journal of Electrochemical Science*, 7(6), pp.5172–5184. doi:https://doi.org/10.1016/s1452-3981(23)19612-x.

[8] Hummers, W.S. and Offeman, R.E. (1958). *Journal of the American Chemical Society*, 80(6), pp.1339–1339. doi:https://doi.org/10.1021/ja01539a017.

[9] Rivera, M., Akashi, L., Rodríguez, L.M., Pérez, P.D., Barboza, C.M., Batista, H. and Champi, A. (2025). *Diamond and Related Materials*, 154, p.112204. doi:https://doi.org/10.1016/j.diamond.2025.112204.

[10] Yu, H., Zhang, B., Bulin, C., Li, R. and Xing, R. (2016). *Scientific Reports*, 6(1). doi:https://doi.org/10.1038/srep36143.

[11] De Silva, K.K.H., Huang, H.-H. ., Joshi, R.K. and Yoshimura, M. (2017). *Carbon*, 119, pp.190–199. doi:https://doi.org/10.1016/j.carbon.2017.04.025.

[12] S. Kumuda, Gandhi, U., Umapathy Mangalanathan and K. Rajanna (2024). *Journal of Materials Science Materials in Electronics*, 35(9). doi:https://doi.org/10.1007/s10854-024-12393-y.

[13] Brusko, V., Artur Khannanov, Aydar Rakhmatullin and Dimiev, A.M. (2024). *Carbon*, 229, pp.119507–119507. doi:https://doi.org/10.1016/j.carbon.2024.119507.

[14] Eigler, S., Dotzer, C., Hof, F., Bauer, W. and Hirsch, A. (2013). *Chemistry - A European Journal*, 19(29), pp.9490–9496. doi:https://doi.org/10.1002/chem.201300387.

[15] Minitha, C.R. and Rajendrakumar, R.T. (2013). *Advanced Materials Research*, 678, pp.56–60. doi:https://doi.org/10.4028/www.scientific.net/amr.678.56.

[16] Pinheiro, T.S., Rondan, W., Malateaux, M.F., Lozada, M., Abê, I.Y., de Assunção, R.M.N., Albertin, K.F. and Champi, A. (2025). *Brazilian Journal of Physics*, 55(4). doi:https://doi.org/10.1007/s13538-025-01781-2.

[17] Lee, A., Yang, K., Anh, Park, C.-H., Seung Hwan Lee, Tae Hoon Lee and Mun Seok Jeong. (2021). 536, pp.147990–147990. doi:https://doi.org/10.1016/j.apsusc.2020.147990.

[18] Feng, J., Ye, Y., Xiao, M., Wu, G. and Ke, Y. (2020). *Chemical Papers*, 74(11), pp.3767–3783. doi:https://doi.org/10.1007/s11696-020-01196-0.

[19] Laura, Bichara, Hernan, Lanus, Silvia and Brandan. (2011). *Journal of Chemistry and Chemical Engineering*, 5(10), pp.936–945.

[20] Guerrero-Contreras, J. and Caballero-Briones, F. (2015). *Materials Chemistry and Physics*, 153, pp.209–220. doi:https://doi.org/10.1016/j.matchemphys.2015.01.005.

[21] Marcano, D.C., Kosynkin, D.V., Berlin, J.M., Sinitskii, A., Sun, Z., Slesarev, A., Alemany, L.B., Lu, W. and Tour, J.M. (2010). *ACS Nano*, 4(8), pp.4806–4814. doi:https://doi.org/10.1021/nn1006368.

[22] Cançado, L.G., Jorio, A., Ferreira, E.H.M., Stavale, F., Achete, C.A., Capaz, R.B., Moutinho, M.V.O., Lombardo, A., Kulmala, T.S. and Ferrari, A.C. (2011). *Nano Letters*, 11(8), pp.3190–3196. doi:https://doi.org/10.1021/nl201432g.

[23] Gagneten, A.M., Rondan, W., Schmuck, J., Reno, U., Regaldo, L., Houre, M., Kladniew, B.R., Rivera, M., Vazquez, J., Acuña, J.J.S. and Champi, A. (2025). *Journal of Applied Phycology*, 37(3), pp.1851–1869. doi:https://doi.org/10.1007/s10811-025-03506-z.